# Security, Privacy and System-Level Resillience of 6G End-to-End System: Hexa-X-II Perspective

Pawani Porambage[1], Diego Lopez[2], Antonio Pastor[2], Bin Han[3], José María Jorquera Valero[4], Manuel Gil Pérez[4], Noelia Pérez Palma[4], Antonio Skarmeta[4], Prajnamaya Dass[5], Stefan Köpsell[5], Sonika Ujjwal[6], Javier José Díaz Rivera[7], Pol Alemany[7], Raul Muñoz[7], Jafar Mohammadi[8], Chaitanya Aggarwal[8], Betul Guvenc Paltun[9], Ferhat Karakoc[9]

[1]VTT Technical Research Centre of Finland, [2]Telefonica I+D, [3]RPTU University of Kaiserslautern, [4]Universidad de Murcia, [5]Barkhausen Institut gGmbH, [6]Oy LM Ericsson Ab, [7]Centre Tecnològic de Telecomunicacions de Catalunya (CTTC/CERCA), [8]Nokia Solutions and Networks GmbH & Co KG, [9]Ericsson Research Türkiye

*Abstract*— The sixth generation (6G) of mobile networks are being developed to overcome limitations in previous generations and meet emerging user demands. As a European project, the Smart Networks and Services Joint Undertaking (SNS JU) 6G Flagship project Hexa-X-II has a leading role for developing technologies and anchoring 6G end-to-end system. This paper summarizes the security, privacy and resilient (SPR) controls identified by Hexa-X-II project and their validation frameworks.

## I. INTRODUCTION

The sixth generation of mobile networks is evolving to meet the unprecedented user requirements that the fifth generation could not fulfill. Many research efforts on 6G are being already undertaken by industry and academia, while standardization entities have also initiated programs for the 6G standardization process. The Hexa-X-II project is significantly contributing to making the 6G vision a reality. Unveiling the design guidelines for the 6G end-to-end system, Hexa-X-II presents a comprehensive view of the 6G E2E system [1]. This includes the infrastructure layer, network function layer, application enablement platform layer, and application layer, all complemented by a set of pervasive functionalities and various forms of connectivity. According to the general view of the 6G E2E system proposed by Hexa-X-II, security and privacy are collectively considered as pervasive functionality that affects all layers as well as the other pervasive functionalities. To cater this, the project brings out the unique identification of relevant technological enablers particularly in the fields of security, privacy, and resilience (SPR). It is important to highlight that these enablers are introduced as SPR controls and intended to address specific threats, providing mechanisms to detect them and to mitigate their impact in the system performance while keeping trustworthiness aspects as a key value. This paper summarizes the SPR controls investigated in Hexa-X-II and the possible evaluation platforms considered for the validation, where further details are presented in project deliverables as referred in [2], [3], [4].

## II. SPR CONTROLS FROM HEXA-X-II

*1. Architectural Enablers* are identified as three main trends conforming the security threat posed by different tendencies towards 6G with respect to the mobile network evolution [2]. These are the network of networks (NoN) compositional pattern, the use of cloud continuum as base infrastrual approach, and the disaggregation of the radio access network (RAN). To address the threats implied by increased complexity of these trends may require the application of the formal security proofs to support correctness analysis and the base mechanisms for confidential usage of computing and networking facilities to asure trust in their deployment.

*2.* The effectiveness and secrecy rate of secret key generation methods based on *Physical Layer Security (PLS)* are dependent on wireless channel. This inidicates the context-awareness in wireless systems leverages environmental factors such as obstacles, movement, and channel conditions. In Hexa-X-II, two implementations were considered that use ultra-wideband Channel State Information (CSI) to demonstrate how these physical parameters influence CSI features [3] [4]. Machine Learning (ML) techniques were employed to classify environments effectively by extracting and evaluating these features, thereby enhancing adaptability in key generation. In addition, Physical Layer Deception (PLD) is an innovative approach to enhance security in wireless communications [4]. It employs a symmetric block encryption algorithm with a randomly selected key from a predefined pool, ensuring that valid ciphertext and plaintext codewords are identical. The key and ciphertext are separately channel-coded, modulated, and power-adapted before transmission. By optimizing coding rates and power mixing ratios, PLD can selectively protect the key component from eavesdropping while exposing the ciphertext. This allows legitimate receivers to decode the full message, while eavesdroppers with poor channel conditions may only decode the ciphertext, misinterpreting it as plaintext due to the encryption codec's closure property. To prevent detection of the ciphering status, PLD can be randomly activated or deactivated. Numerical simulations demonstrate that PLD can outperform classical physical layer security approaches in terms of leakage-failure probability, particularly when the eavesdropping channel is of good quality.

*3.* Hexa-X-II also considers the treats implied by the pervasive use of *Artificial Intelligence (AI)/ML* and the enablers to enhance their trustworthiness at all levels [2]. The use of AI/ML in 6G comprises networks optimization and automation, and security protection against various attacks. The project studies the trustworthiness of AI/ML for 6G by focusing on security, privacy and explainability aspects, and identifies solutions such as Federated Learning (FL) and explainable AI (XAI) [3] [4]. Nevertheless, FL systems are susceptible to security and privacy attacks due to their distributed and open structure. An adversary could poison the training process to forcibly drift the model away from the optimal point or to a point targeted by the attacker. Anomaly detection algorithms (e.g., by utilizing XAI), adaptive regularization methods, norm clipping, and weak Differential Privacy have been proposed to enhance the resilience of FL setups. However, use of these methods can be challenging if both robustness and privacy are needed simultaneously [4].

4. *Joint Communication and Sensing (JCAS)* presents unique security and privacy challenges due to the inclusion of sensitive sensing data. Within Hexa-X-II, the emerging JCAS architecture has been analyzed, revealing a significant need for network functions that address sensing policies, control, and transparency management (SPCTM) [3] [4]. To incorporate SPCTM into the core network, interfaces with other network functions have been designed accordingly. The enhanced JCAS architecture has been evaluated using STRIDE and LINDDUN models to identify various threats, and suggestive measures have been proposed for mitigating these threats [4].

5. Hexa-X-II presents a Level of Trust Assessment Function (LoTAF) to guarantee security aspects for Cloud Continuum scenarios. LoTAF is a neutral, bidirectional, and intelligent service supporting *trustors* (users) in making informed decisions and provisioning *trustees* (network providers) insights into service compliance regarding 6G security requirements. Hence, LoTAF streamlines the trustworthiness assessment of network services before and after they are instantiated, adhering to well-known standards (e.g., ITU-T Y.3057 and Service Assurance for IBN architecture). Concerning the LoTAF functionalities, the key activities are split into two main phases: i) semantic understanding, mapping, and information representation to handle knowledge graphs and agreements in terms of trust between parties; and ii) monitoring of service assurance to check deviations on agreed trust requirements and apply rewards or punishments on an initial LoT. All in all, LoTAF paves the way to enhance security and trustworthiness of 6G network services endowing users and providers with valuable insights and decision-making support.

6. *Quantum resisitive cryptography*: The integration of post-quantum cryptography (PQC) primitives into existing software frameworks is advancing, with libraries implementation, helping to harmonizing PQC with established security protocols such as TLS. As the first mobile network architecture designed with these new cryptographic considerations in mind, the 6G network should not only integrate PQC but also evaluate the consequences of the cryptographic shift on network structure, operations, and efficiency. The transition to quantum-resistant cryptography is also exploring the synergy between PQC and quantum key distribution (QKD), aiming to support adaptative key distribution, necessary for the varied operational demands of telecommunications infrastructure for long transition periods. Current experiments in HEXA-II are focused on impact in current practices for protocol security, especially in terms of availability and, performance.

7. *Distributed Ledger Technologies (DLTs)* provide a robust alternative by enabling decentralized management and immutable recording of network configurations, enhancing security against tampering and unauthorized changes. To this end, Hexa-X-II investigates the use of DLT [1] a private, permissioned ledger where topology changes are recorded as transactions ensuring that only those authorized stakeholders are able to make changes [4]. With that, securely storing and sharing network topology information for network management decisions contributes to achieving a secure and privacy-based multi-stakeholder scenario.

III. EVALUATION PLATFORMS

As persistent security and privacy are paramount in the design of the 6G E2E system, Hexa-X-II considers the integration of comprehensive security frameworks, including quantum-resistant cryptography, trustworthy AI, and confidential network deployment, ensures a secure and trustworthy environment for 6G services [4]. For example, SecDevOps mechanisms in Hexa-X-II systematically quantify service privacy levels using a declarative, service-specific privacy manifest. This manifest, detailing privacy service definitions, is analyzed with predefined metrics and algorithms. The privacy score, derived from both developer-provided and automatically assessed information, comprehensively evaluates data processing, storage, and communication. The architecture supports continuous privacy monitoring and updates, promoting ongoing refinement of privacy practices through DevPrivOps. Moreover, the architecture enhances privacy and trust by integrating several key components. The Management API uses Threat Risk Assessor (TRA) output for privacy quantification, while the Zero-Trust Security and Identity Management component performs continuous trust evaluation of network functions and entities. AI algorithms assist in trust evaluation and provide proactive notifications about trustworthiness. The Security Orchestrator employs AI to manage function chaining and orchestration, ensuring quality, performance, and security across network segments and domains. This orchestrator considers various network parameters and collaborates with other managers to implement countermeasures, thereby ensuring the resilience and reliability of 6G network services. Additionally, the work emphasizes practical outcomes, such as the DevSecOps pipeline and the use of what-if scenarios in a Network Digital Twin (NDT) to evaluate the impact of different measures. Hexa-X-II aims to integrate these results and other advancements from the SNS projects 6G research.

ACKNOWLEDGMENT

This work is funded by the EC through the Hexa-X-II (101095759) project.